\documentclass[iop]{emulateapj}

\newcommand{\sinc}{{\rm sinc}}

\usepackage{color}
\usepackage{graphicx}
\usepackage{epsfig}

\usepackage{natbib}

\usepackage{amsmath}

\def\simlt{\lower.5ex\hbox{$\; \buildrel < \over \sim \;$}}

\def\simgt{\lower.5ex\hbox{$\; \buildrel > \over \sim \;$}}

\begin{document}

\title{Precision Near Infrared Photometry For Exoplanet Transit Observations - I : Ensemble Spot Photometry for An All-Sky Survey}
\author{C. Clanton$^{1,2}$, C. Beichman$^{3,4}$, G. Vasisht$^3$, R. Smith$^5$ and B. S. Gaudi$^2$}
\affil{1) Caltech Optical Observatories, California Institute of Technology, Pasadena CA 91125}
\affil{2) Department of Astronomy, Ohio State University, Columbus, Ohio 43210}
\affil{3) Jet Propulsion Laboratory, California Institute of Technology, Pasadena CA 91107}
\affil{4) NASA Exoplanet Science Institute, California Institute of Technology, Pasadena CA 91125}
\affil{5) California Institute of Technology, Pasadena CA 91125}

\affil{chas@pop.jpl.nasa.gov,rsmith@astro.caltech.edu, gv@jpl.nasa.gov}

\begin{abstract}

Near-IR observations  are important for the detection and characterization of exoplanets using the transit technique, either in surveys of large numbers of stars or for follow-up spectroscopic observations of individual planets. In a controlled laboratory experiment, we imaged $\sim 10^4$ critically sampled spots onto an Teledyne Hawaii-2RG (H2RG) detector to emulate an idealized star-field. We obtained time-series photometry of up to $\simeq 24$ hr duration for ensembles of  $\sim 10^3$ pseudo-stars. After rejecting correlated temporal noise caused by various disturbances, we measured a photometric performance of $<$50 ppm-hr$^{-1/2}$  limited only by the incident photon rate. After several hours we achieve a photon-noise limited precision level of $10\sim20$ ppm after averaging many independent measurements. We conclude that IR detectors such as the H2RG can make the precision measurements needed to detect the transits of terrestrial planets or detect faint atomic or molecular spectral features in the atmospheres of transiting extrasolar planets.

\end{abstract}


\section{Introduction}

The dramatic results of the Kepler and COROT missions \citep{Borucki2011, Baglin2009} as well as earlier ground-based surveys have demonstrated the power of the transit method for discovering and characterizing other planetary systems. The combination of a planetary radius from a transit measurement with a mass from a radial velocity measurement yields a planet's density and thus clues to its composition and evolutionary history. Optical and infrared observations can determine a planet's albedo and equilibrium temperature and even probe global atmospheric circulation patterns. Spectroscopic transit observations have identified atomic and molecular species, including H$_2$O, CO$_2$, and CH$_4$ in the atmospheres of planets orbiting nearby bright stars. 

The critical instrumental requirement for all these observations is highly precise differential photometry since the transit of a Jupiter in front of a solar type star (1~R$_\odot$) causes only a 1\% decrease in brightness while the transit of an Earth in front of the same star causes only a 0.0084\% dip (where 0.01\% = 100 ppm). Wide-field surveys from the ground are typically limited in precision by residual atmospheric effects to 2,000$\sim$4,000 ppm, e.g. Super-WASP \citep{Pollacco2006} and HATNet \citep{Bakos2009}. Higher precision measurements, $<$1,000 ppm, has been demonstrated with concentrated observations of individual stars \citep{Irwin2010, Johnson2009}. From space, however, much higher levels of precision can be achieved for hundreds of thousands of objects simultaneously. The Kepler and COROT missions have demonstrated photometric precision in the visible using CCD arrays as low as 15 ppm, limited in many cases by stellar variability itself \citep{Jenkins2011b}. Spitzer and HST photometry and spectroscopy using earlier generations of near- and mid-IR detectors have demonstrated precision around 100 ppm \citep{Knutson2007, Swain2009, Deming2006}.

There are many advantages to carrying out transit observations in the near-IR compared to the visible band: the photon fluxes from K and M stars are higher in the near-IR, up to a factor of 50-100 for the coolest M stars; total photon fluxes from hotter F and G stars are factor of $\sim$2 higher in a wavelength band that extends out to 1.6 $\micron$\ rather than in a visible band that cuts off at the 1 $\micron$ CCD limit; starspot noise is reduced compared to visible wavelengths due to the lower contrast between areas with typical spot and photospheric temperatures  \citep{Frasca2009}; and the effects of limb-darkening are greatly reduced, simplifying extraction of planetary parameters from the observations \citep{Desert2011}.

\begin{figure*}
\epsscale{1}\plotone{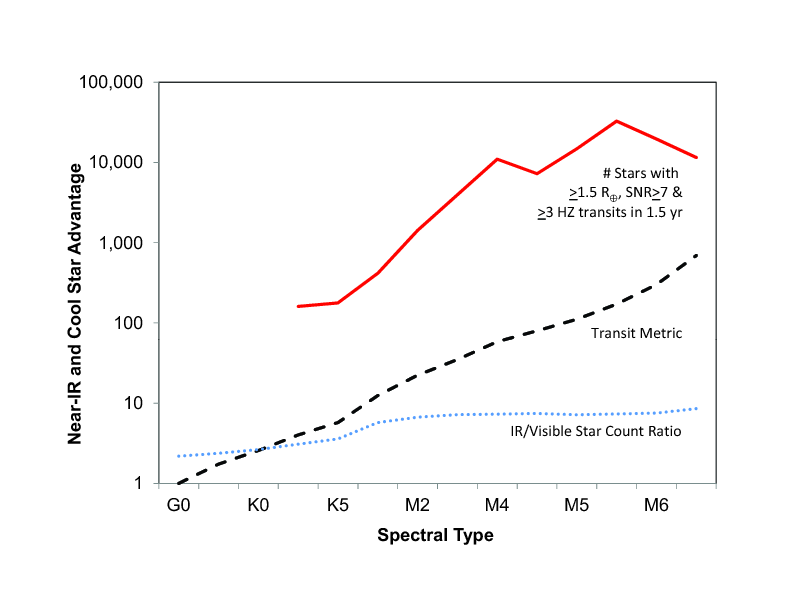} \caption{The 10-1,000-fold {\it Cool Star  Advantage} of lower luminosity K and M stars relative to a G0 star for detecting habitable zone (HZ) transits is illustrated by a geometrical metric which takes into account the increased alignment probability, the increased frequency of transits, and the increased transit depth as a function of spectral type (black dashed line; \citet{Gould2003}). The {\it Near-IR Advantage} (blue dotted line) is the ratio of the number of stars detectable at a given signal to noise ratio (SNR) in  a broad J$_W$ passband (0.66-1.70 $\mu$m) with H2RG detectors compared to the number detectable with CCDs working in the visible (0.6-1.0 $\mu$m).  A near-IR survey could detect twice as many planets orbiting F and G stars and between 3-7 times more planets around K and M stars. Finally, a Monte Carlo simulation discussed in $\S$5 shows that $\sim$115,000 stars in an all-sky survey lasting 1.5 years would be bright enough for the detection of an 1.5 R$_\oplus$ planet located in the habitable zone (HZ) of its host star (solid red curve). In this context a detection requires the observation of at least 3 transits and a mission total SNR$\ge$7. \label{NIR}}
\end{figure*}

The detection of transiting K and M stars is of great scientific interest since the smaller radii of late type stars enhances the signals from Earth-sized planets \citep{Gould2003}. The transit of an Earth in front of an M8 star produces a 1\% brightness dip, 100 times larger than for a G2 star. Additionally, the location of the habitable zone (HZ) where a planet's temperature might allow for the existence of liquid water, moves in closer to the star, $a_{HZ} \propto L_\star^{1/2} \propto M_\star^{\alpha/2}$ with $\alpha \approx 3.4$, increasing both the probability of a geometrical alignment and the frequency of transits (a few days or weeks compared to a year for a G2 star). Initial results from the Kepler mission suggest that low mass stars have an increased incidence of low mass planets \citep{Howard2011} further enhancing the importance of later spectral types. Figure~\ref{NIR} highlights the increased detectability of transiting planets in the habitable zones of late type stars. The recognition that cool stars offer an excellent opportunity for detecting HZ planets has lead to ground-based searches directed at M-dwarfs \citep{Charb2009} and generated considerable interest in improving near-IR Doppler precision \citep{Reiners2010, Bean2010, Plavchan2011}.

The large signal from a planet transiting a cool star also enhances the spectroscopic potential of these systems. As pointed out by \citet{Deming2009}, transit observations using the James Webb Space Telescope (JWST) could identify numerous atomic and molecular species in the atmosphere of a $\sim$2 R$_\oplus$ planet transiting a M5 star. Future missions have also been proposed to exploit the potential of spectroscopy to characterize exoplanet atmospheres. A critical need for successful transit spectroscopy are planets orbiting {\it nearby, bright stars} so that the necessary precision can be achieved in a reasonable time. Kepler and COROT targets are too faint for this purpose and planets discovered by ground-based transit surveys are too few in number and limited to relatively large radii due to the constraints of observing through the atmosphere.

Because nearby bright stars are distributed isotropically, there is interest in conducting an extremely wide field-of-view panoramic survey using a low cost orbital platform \citep{Ricker2010, Deming2009, Beichman2009}. Such a survey would target 1-2 million stars which would be at least 5 magnitudes brighter than the Kepler stars. Such a survey could be conducted with multiple 12.5 cm apertures, a factor 10 smaller than Kepler's 1 m telescope, operating either in the visible using CCD focal planes \citep{Ricker2010, Deming2009, Claudi2010} or in the near infrared \citep{Beichman2009}. Shifting the observational band from the visible (0.6-1.0 $\mu$m) into the near-IR (0.66-1.65 $\mu$m)\footnote{The short wavelength cutoff is selected to exclude H$\alpha$ emission which can be variable in cool stars. The long wavelength cutoff is tailored to exclude small temporal or spatial variations in detector QE at the limit of its responsivity.} for otherwise identical survey facilities results in a 3 to 10-fold increase in the numbers of cool stars (K5-M5) bright enough for the detection of $\sim1$ R$_E$ planets (dashed line in Figure~\ref{NIR}). A Monte Carlo simulation $(\S$5) shows that an all-sky survey of 2 million stars would include $\sim 150,000$ stars  bright enough to allow the detection of planets as small as 1.5 R$_\oplus$ located in the habitable zone (HZ) of its host star with a mission total signal to noise ratio of 7 or higher. The vast majority of these stars have spectral types later than M0 and are thus preferentially detected in the near-IR.

The increased near-IR yield over the visible comes from the increased signal-to-noise ratio (SNR) in the near-IR due to stronger photospheric emission in the  near-IR  \citep{haus99} and/or broader measurement bandwidth.  The star counts above a given flux threshold increase  as the photon luminosity in a bandpass ($L_\lambda\Delta\lambda$) to the 3/2 power.  The number of stars above a given SNR threshold goes as $N \propto SNR^{-3} (L_\lambda\Delta\lambda)^{3/2}$, and thus  by increasing  $L_\lambda\Delta\lambda$,  the number of stars above a fixed SNR increases as $(L_\lambda\Delta\lambda)^{3/2}$. Figure \ref{NIR} also shows that the near-IR measurements do  as well or better than a comparable visible light mission for hotter F, G, K stars for the reasons mentioned above. 

It is with these scientific goals in mind --- an all sky survey to identify transiting planets with an emphasis on cool K and M stars and follow-up spectroscopic observations using stable space telescopes --- that we have initiated a series of laboratory experiments designed to probe the ultimate limits of precision photometry using near-IR detectors. We report on the photometric properties of Teledyne's widely-used, infrared H2RG detector\footnote{H2RG or HAWAII-2RG: HgCdTe Astronomical Wide Area Infrared Imager}, which offers near unity quantum efficiency in the near infrared \citep{Beletic2008}. In this paper we describe laboratory experiments (\S~2) to demonstrate the technical readiness of space-borne, transit experiments using existing near-IR detectors. In a future paper we will describe the results of simulated differential spectroscopy planned for the JWST and proposed for other missions.

\subsection{Photometry with Infrared Focal Planes} 

There is broad interest in the performance characteristics of the next generation of IR detectors. H2RGs will be flown on multiple {\it James Webb Space Telescope (JWST)} instruments \citep{Clampin2010, Deming2009} and may also form the basis for dedicated exoplanet observatories such as the candidate Explorer mission FINESSE and the proposed European {\it Exoplanet Characterization Observatory} (ECHO) \citep{Tinetti2011, Tessenyi2012}. Although various investigators have measured relevant device-level properties such as the subpixel non-uniformity \citep{Barron2007, Hardy2008}, the count-rate dependent non-linearity or reciprocity failure \citep{Bies2011}, and image persistence \citep{Smith2008}, there has not been a systematic study of the limits to high precision time-series measurements.

Our goal is to assess the precision possible with the current generation of NIR devices, and whether it is possible to achieve IR photometry that is significantly better than what is being achieved from space today. The current state-of-the-art is defined by the {\it Spitzer} and {\it Hubble} Space Telescopes. Observations to determine atmospheric composition by detecting variations in the apparent size of the planet as a function of wavelength using filter photometry or low dispersion spectra have had notable successes, but also some controversy. Claims have been made about detection of water in Spitzer transit and secondary eclipse photometry \citep{Barman2007, Barman2008, Tinetti2007} as well as counter-claims \citep{Desert2009} --- sometimes using the same data set. Similarly, the detection of molecular features due to methane, water and carbon dioxide in NICMOS spectrophotometry is controversial \citep{Swain2009, Gibson2011}. Observations of the hot Neptune GJ 436b have also led to controversial claims regarding the presence or absence of methane in the atmosphere \citep{Stevenson2010, Beaulieu2011, Knutson2011}. To a great extent it is the treatment of small, systematic detector effects that lie at the heart of these debates. 

The results to date suggest that the instrumental precision of {\it Spitzer} IRAC photometry and HST NICMOS spectroscopy
is $\sim 100$ ppm, although recent improvements in Spitzer's pointing stability have improved that IRAC's photometric precision to $<$50 ppm \citep{2012ApJ...751L..28D}. A combination of factors, including detector non-linearities, poor sampling of the Point Spread function (PSF), short term wobbles and long term drifts in telescope pointing  change the measured signal from the star by several percent over the course of an hour and make the extraction of millimag features from IRAC light curves very challenging. Similarly, {\it Hubble's} low Earth orbit introduce systematic effects on the transit timescale \citep{Swain2008, Swain2009, Gibson2011, Berta2011}.

\begin{figure}
\epsscale{1}\plotone{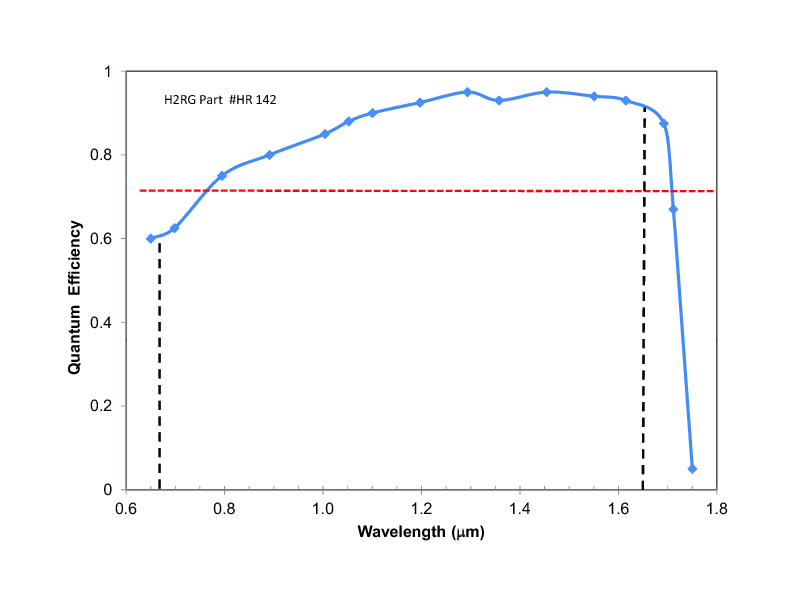} \caption{The substrate-removed H2RG \citep{Beletic2008} detector used in these experiments shows high quantum efficiency, $>$85\%, over a broad visible and near-IR band out to 1.7$\micron$ at an operating temperature of 145K.  The horizontal red line show a typical mission requirement which should be readily achievable with existing technology. Vertical dashed lines show a nominal system passband designed to limit variability due to stellar H$\alpha$ emission at the short wavelength end ($\lambda>0.66\, \mu$m) and temperature induced detector QE variations at the long wavelength end ($\lambda<1.65\, \mu$m).\label{QE}}
\end{figure}

\begin{figure*}
\epsscale{1}\plotone{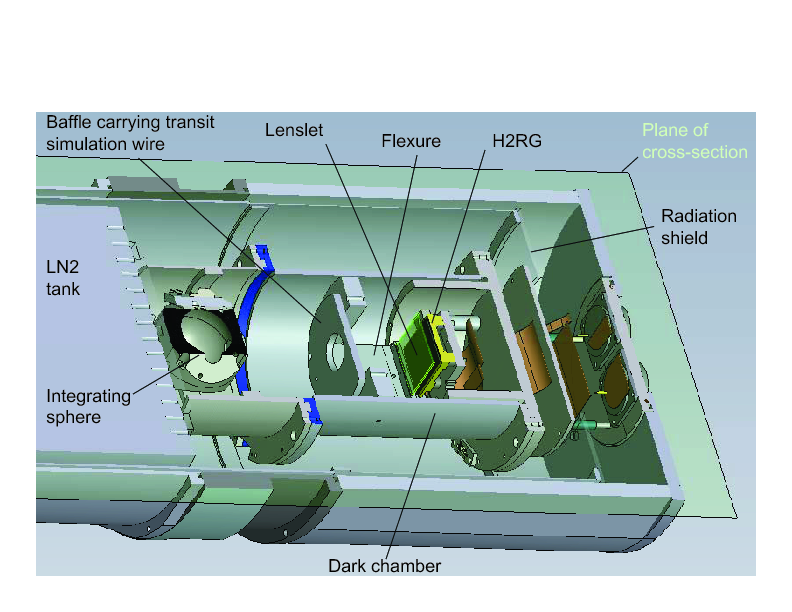} \caption{The dewar includes an LED illuminating an integrating sphere, a lenslet array, and the H2RG detector. There is no external source of illumination into the dewar.\label{layout}}
\end{figure*}

\subsection{Lightcurve Measurements}

When measuring transit lightcurves, the timescale of relevance is of order the transit duration, $t_d$. An Earth-Sun transit produces a characteristic signal of depth $84 \left (\frac{R_p}{R_\oplus}\frac{R_\odot}{R_*}\right )^2$ ppm with a maximum duration of 13 hr. More generally, the maximum duration with a star-planet separation of $a$ (in AU) is $t_d \simeq 13  (a/AU)^{1/2} (M_*/M_{\odot})^{1/2}$ hr assuming $R_* \propto M_*$, which is a good approximation for the lower
main sequence. For a solar type star transits by planets with separations of $a \sim $ 0.01, 0.1 and 1 AU have durations of $\sim$ 1, 4 and 13 hr, respectively. 
Three distinct sources of noise establish the detection threshold of a transit observation over the measurement time 
$t$: 1) the noise from photon arrival statistics with a standard deviation $\sigma_N$; 2) noise from intrinsic stellar variability, $\sigma_S$; and 3) the instrument measurement noise, $\sigma_I$. The total photometric noise $\sigma_T$ is given by 

\begin{equation}
\sigma_T = (\sigma_N^2+\sigma_S^2+\sigma_I^2)^{1/2}.
\end{equation}

\noindent In Kepler parlance $\sigma_T$ is called the combined differential photometric precision, or CDPP \citep{Jenkins2011b}.

A few general remarks are important to consider in designing a laboratory experiment to test the limits of detector performance in this application. First,  
the photometric noise $\sigma_T$ is  inherently a differential measurement, because only brightness differences in and out of transit are important. Thus, neither absolute photometric calibration nor instrumental or stellar variability changes with timescales much longer than a few times $t_d$ are relevant to the measurement. 
Second, over the long term, i.e., several transits, the stellar noise, $\sigma_S$, should average down to 10 ppm for quiescent stars \citep{Basri2010, Ciardi2011}. The Sun and some of its G-dwarf cousins are relatively quiet in this respect \citep{Basri2010, Jenkins2011a}  with $10^3$ ppm variation on timescales of a rotation period or longer but with relatively little variability on the 10-13 hr timescale of an Earth transit. While some stars are as quiet as the Sun, 10 ppm, on timescales of a few hours, Kepler has shown that {\it at visible wavelengths} the average solar-type star is somewhat noisier than the Sun, 10$\sim$30 ppm \citep{Jenkins2011b}. The vast majority of field M dwarfs observed by Kepler have short-term variability (0.5 to 12 hour) less than 1000 ppm and many with $<$100 ppm \citep{Ciardi2011, Bryden2011}. However, it is important to note that variability due to starspots will be a factor of 2-3 smaller in the near-IR than in the visible due to reduced contrast in the Rayleigh-Jeans portion of the spectrum \citep{Frasca2009}. Finally, in a properly designed system, $\sigma_I \simeq \sigma_S$, so that little is to be gained by reducing any one of the noise components to be significantly smaller than the total. The goal of our experiments is to identify methods of reducing $\sigma_I$ so that it is comparable to the other irreducible terms (10-20 ppm). Quantifying the magnitude of these effects and demonstrating an ability to reject them as part of a total error budget is critical to designing a particular experiment. Major contributions to $\sigma_I$ in transit experiments are pixel-scale detector inhomogeneities (``flat fielding errors'') convolved with  pointing changes, responsivity drifts with temperature and other environmental effects, and detector non-linearities such as latent image generation.

In $\S$2, we describe our experiments and data acquisition. In $\S$3, we discuss our methods and describe the custom data analysis pipeline that was written to analyze the large quantities of data that were produced by the experiments (up to 200 Gbytes/day). Results are presented $\S$4.

\section{Description of the Experiment \label{EXPT} }
The experimental setup described in this section was used to  investigate three major sources of degradation in high precision differential photometry: 1) long term ($\sim$24 hr) drifts, including the response of the detector and  warm electronics, variation in brightness of the illuminating LED and environmental factors such as dewar temperature; 2) effects of simulated pointing drifts and offsets; and 3) the effects of changes in detector temperature. The magnitude of these effects and the development of mitigation techniques are of direct relevance to future space missions  where the expense of controlling  parameters such as detector temperature and spacecraft pointing must be traded against the exacting demands of transit measurements.

	A 1.7 $\mu$m cutoff HgCdTe hybrid CMOS detector with 2044x2044 imaging pixels on 18 $\mu$m\ pitch, Teledyne H2RG-103, was mounted on a temperature-controlled Molybdenum plate inside a sealed chamber refrigerated to 77K by a liquid Nitrogen dewar with $>$20 hr hold time. The wavelength dependence of the detector responsivity is shown in Figure~\ref{QE}. The dewar was filled $\sim$1 hour prior to each experiment to allow for thermal stabilization and thereafter was automatically refilled every 20 hours from a 160 liter storage dewar so that the detector never warmed up. 

	The dark chamber was evacuated through a 2.5 mm diameter serpentine black-painted tube to block entry of thermal emission from warmer parts of the dewar. Light leaks and thermal background were undetectable even in 4 hour exposures against a 0.001 e$^-$s$^{-1}$ pixel$^{-1}$ dark current floor obtained by cooling the detector to 110K.  At 140K, this detector exhibits median dark current $<$0.02 e$^-$s$^{-1}$ pixel$^{-1}$ which is negligible compared to all signals in this experiment.  The $\sim20$  e$^-$/pixel read noise (for correlated double sampling) from the 150 pixels in the signal and background apertures combined contributes less than 1/3 of the total variance in the signal compared with the signal per pseudo star of 300,000 e$^-$ per 1.2 sec integration. 

	The layout of the dark chamber is shown in Figure~\ref{layout}. Two LEDs within the dark chamber (wavelengths 0.890 and 1.30  $\mu$m) are driven by precision current sources external to the dewar. They shine concurrently into a 2\arcsec\ integrating sphere with a finely textured gold interior and a precision 1 mm exit aperture. The whole illumination system is housed behind a baffle so that only light from the exit port reaches the detector 139 mm away. A second baffle half way along the light path blocks rays scattering off the sidewalls and shades the lenslet supports.  A 40 mm by 40 mm fused silica lenslet array made by SUSS Microptics SA (Switzerland), with 300 $\mu$m lenslet pitch and 4.4$\pm$0.4 mm focal length was supported 5 mm from the detector surface to create slightly defocussed image of the 1 mm aperture per lenslet. The precisely rectilinear grid of spots was found to have a 17.29 pixel pitch, as expected when projection effects are taken into account.

\begin{figure}
\epsscale{1}\plotone{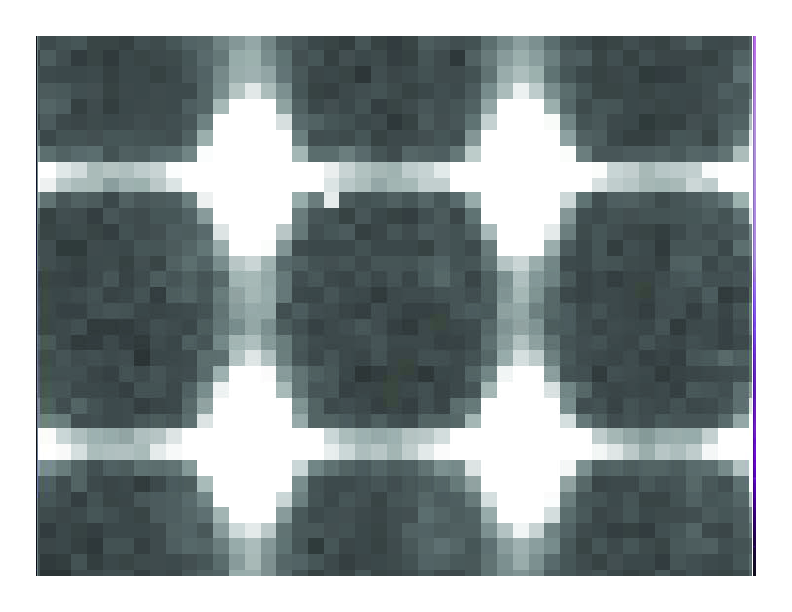} \caption{Four of the $\sim$10$^4$ spots with the scale zoomed to show the individual spots plus diffraction spikes.\label{spots}} 
\end{figure}

\begin{figure}
\epsscale{1}\includegraphics[scale=.35,angle=90]{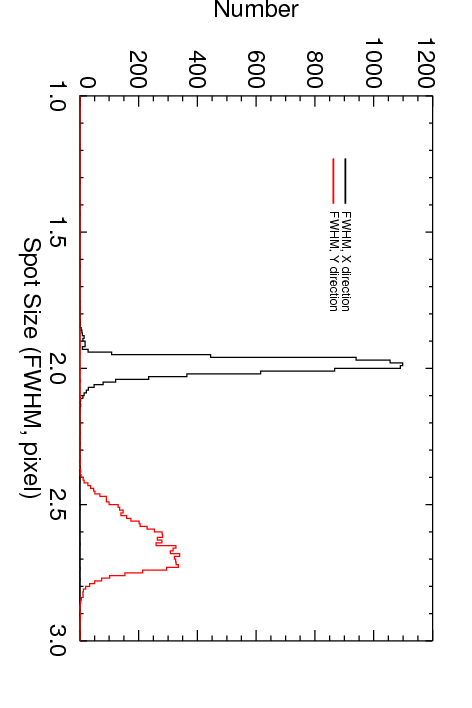} \caption{The spot images generated by the lenslet array are uniform and slightly elongated. The distribution of spot sizes (Full Width at Half Maximum, FWHM) is narrow and centered around 2.0 pixels in the X-direction and 2.5 pixels in the Y-direction.\label{FWHM}}
\end{figure}

	The Full Width Half Maximum (FWHM) of each spot was typically 2.3-2.5 pixels with low amplitude diffraction spikes extending in the vertical and horizontal directions (Figures~\ref{spots} and \ref{FWHM}), as expected for a lenslet array with this pitch.  The maximum flux per spot was $\sim$60 ke$^-$ s$^{-1}$ corresponding to the expected count rate from a 9th magnitude star observed using a camera with a 12.5 cm aperture in a very broad photometric band extending from 0.66 to 1.65 $\mu$m.

\begin{figure*}
\epsscale{1}\plotone{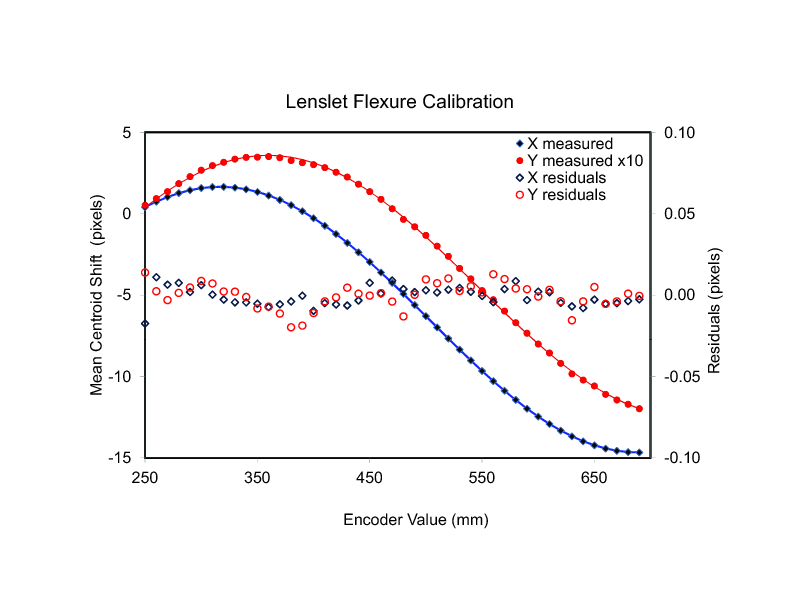} \caption{As the dewar is rotated around its horizontal axis, the average position of the spots shifts by $\pm$8 pixels along the rows (X-direction).  A simple sinusoidal model provides a good fit to the data with residuals at the level of a few hundredths of a pixel. Note that the shifts along columns (Y-direction) are plotted on a 10-times exaggerated scale.\label{rotation}} 
\end{figure*}

	A simple but precise way to move the lenslet array was utilized to emulate small pointing motions. Smooth, reproducible motion with fine resolution was produced by changing the gravity vector acting on the lenslet support formed by two thin, cantilevered Aluminum sheets, which flex predominantly in one direction. The cylindrical vacuum vessel lies horizontally in a wooden cradle with felt padding, allowing the orientation of the flexures to be changed smoothly with respect to gravity by manually rotating the vacuum vessel about the optical axis.  A metal measuring tape attached to the vacuum vessel with a fixed pointer mounted on the cradle, serves as an angle indicator. The lenslet position is self encoding: while the position of any given spot can be can vary due to sampling effects by as much as 180 nm depending on registration with the pixel grid, this effect averages away rapidly and averages of a thousand spots exhibit 2 nm rms temporal stability, and comparable coherence under translation when comparing one half of the spots to the other. Figure~\ref{rotation} shows that the lenslet position varies as the sine of the indicated angle to remarkable accuracy for a hand driven actuator.  Motion up to $\pm$8.2 pixels is produced at an angle 5.2$^o$ from the row direction. The 0.005 pixel (90 nm) rms scatter about a sine fit is dominated by the 0.5 degree accuracy with which the dewar rotation angle can be read. This resolution is more than adequate for pointing control, while the 2 nm temporal stability far exceeds requirements.

	Vibration might be a concern with a flexible lenslet support. The resonant frequency of a mass suspended by a spring is $\omega= \sqrt{g/d} $, where $g$ is acceleration due to gravity and $d$ is displacement due to gravity.  Measured peak displacement is 148 $\mu$m, so the resonant frequency is expected to be 41 Hz. Indeed, the only resonance detected, when applying an impulse to the dewar while reading a small window at $\sim$1200 Hz, was at 41 Hz.  No power was detectable at this or any other frequency when the system was not excited. Nonetheless thick foam padding was placed between the dewar cradle and bench to increase the attenuation seventy fold at 41 Hz assuming moderately damped simple harmonic motion.

	The detector was read out using a 32 channel Astronomical Research Cameras controller with a 11.44 $\mu$s clock time, including  3 $\mu$s integration time and 1.28 $\mu$s settling time.  Correlated double sampling (CDS) was executed using conventional readout waveforms. As these waveforms were not optimized for such short exposures there was an exposure duty cycle penalty which will be reduced in future experiments. When idle, identical clocking waveforms were executed to maintain a stable self-heating pattern and to avoid abnormal charge accumulation. One full frame scan was assigned to line by line reset, rastering through all pixels with identical timing to the readout so that the self heating footprint during reset matched the read scans. This was immediately followed by a read scan to measure the post-reset offset, then an identical read scan to measure integrated signal plus offset. The difference of the two samples for each pixel is the signal. The exposure time is the delay plus one frame scan time, while the exposure periodicity is the delay plus three frame scan times.  The exposure duty cycle is their ratio, 33\%.  To improve the duty cycle, while avoiding some cosmetic defects, the frame scan time was reduced from  1652 msec  for a full frame to 1221 msec, by skipping 400 lines and then reading a band 1513 lines high.

	A typical observing sequence consisted of filling the dewar, setting the LED intensity, and starting a steady cadence of one 1.2 sec exposures every 3.6 sec, repeated continuously for periods as long as the 20 hr LN2 autofill interval.  This experiment was repeated with the same timing for a range of illumination intensities. In one test, the intensity was deliberately modulated to measure the effectiveness of normalization by the ensemble. In other cases,  the dewar was rotated to move the images, simulating pointing changes. Other disturbances such as temperature changes were introduced to assess their importance and our ability to correct for them.

\section{The Analysis: Method and Data Pipeline \label{analysis}}

Raw image data for each experiment were run through a data analysis pipeline that was written specifically for the experimental layout described in $\S$2. Individual steps of the pipeline are summarized below.

\begin{itemize}

\item	Define a region of interest on the array selecting an area of high cosmetic quality for both the detector and lenslet arrays, roughly 30 by 30 spots in size. For analysis we typically used 1,000 spots avoiding bad pixels. Channel-to-channel variations in the warm electronics readouts can introduce small measurement perturbations which can be reduced by rigorous design of the warm electronics and standard destriping techniques. Since for these experiments we wanted to concentrate on the fundamental detector performance issues, we avoided pixels spanning channel boundaries. 

\item 	Construct a master dark frame from 200 continuously acquired dark frames and subtract from each image. No flat field correction was applied to the data since the analysis steps described below provided the equivalent of a highly localized flat field for each spot.

\item	Determine centroids for each star for use as the center for aperture photometry. The IDL routine FIND\footnote{IDL Astronomy User's Library} locates sources by fitting a 2-D Gaussian, yielding the centroids and full width at half maximums (FWHMs) in orthogonal directions for each star. Depending on the experiment, we developed two different ways of placing apertures. In experiments where the stars are stationary with respect to the focal plane, i.e. with no intentional pointing disturbances, we used stationary apertures which remained at a fixed position throughout the photometric sequence. To account for any drifts that may occur, as well as any estimation errors that may arise due to quantization effects of a spot lying on a pixel boundary, the aperture centers are computed as follows: the first few ($\sim 5$) frames of the sequence are averaged together and then FIND is used to locate the spots in this averaged frame. The same is done with frames from the middle and end of a sequence. This produces centroids for each spot which are then averaged together to get the final value of the aperture center. Typical image motions from the beginning to the end of the photometric sequence are $\sim$ 0.03 pixels and $\sim$ 0.004 pixels for X and Y centroids, respectively. In experiments with deliberate pointing disturbances, the apertures are allowed to track the star and the moving centroids were recomputed every frame. 

\item	Calculate fluxes, $l(t)$, using the IDL routine APER\footnotemark[3] with a 3 pixel radius aperture which captures $\sim$ 95\% of the point spread function's encircled energy. The signal contained within a background annulus of inner and outer radii of 5 and 8 pixels was subtracted to remove background counts.

\end{itemize}

\begin{figure}
\epsscale{1}\plotone{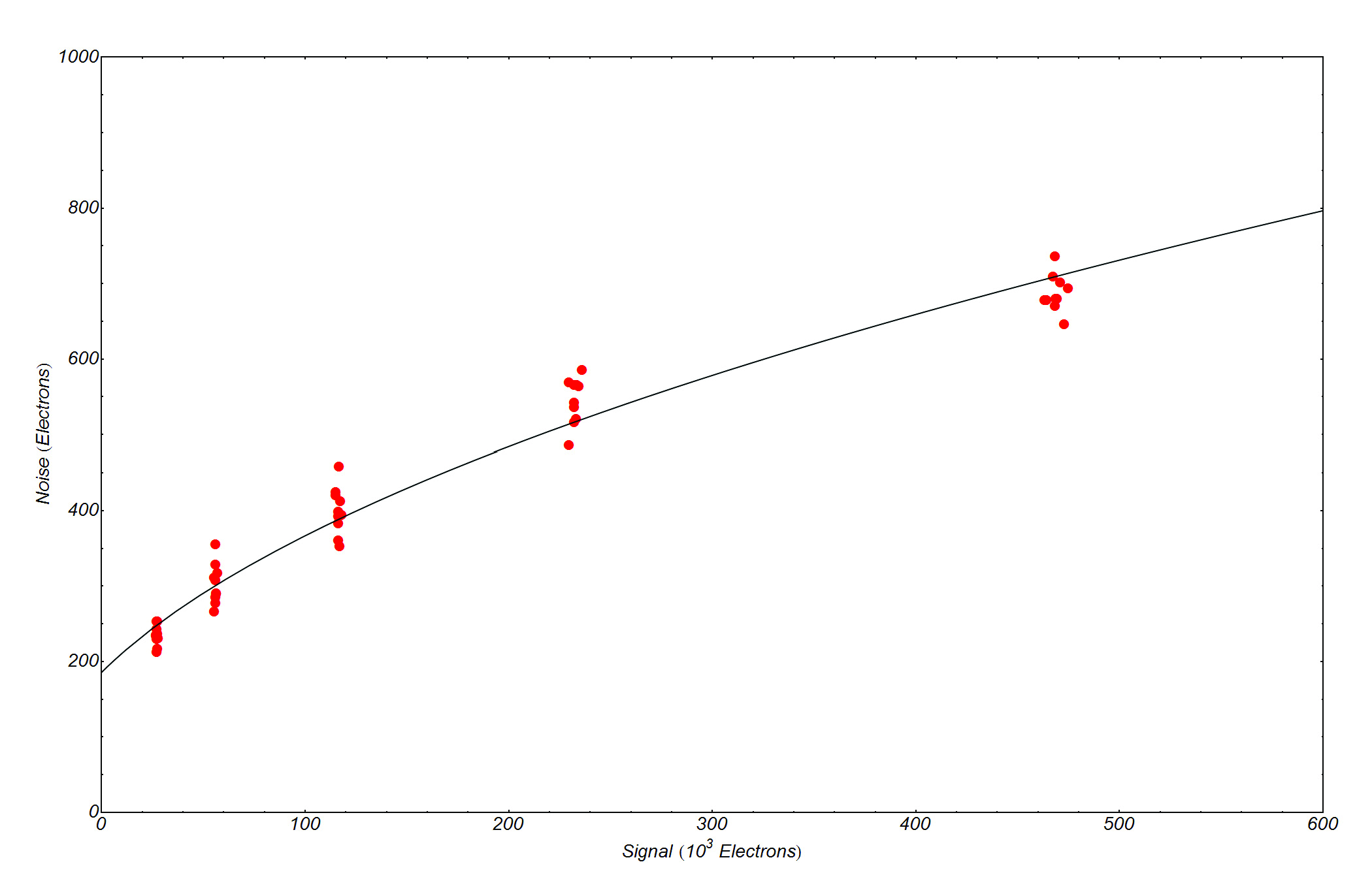} \caption{Measured noise in 10 photometric apertures is shown as a function of signal brightness along with a a simple model combining read noise and photon shot noise (Eqn.~2). The measurements are photon shot noise limited.\label{shot}} 
\end{figure}

We first use  aperture photometry of 10 spots to demonstrate  that the detector performance is limited by photon shot-noise. Measurements  at 5 different brightness levels were fitted to a noise model (Eqn~2) relating the measured noise to the signal intensity and read noise (RN) in $N_{pix}=150$ pixels (signal aperture plus background annulus):

\begin{equation}
\mathrm{Noise}(e^-)= \sqrt{\mathrm{Signal}(e^-)+\mathrm{(RN)}^2\ N_{pix}}
\end{equation}

\noindent where the Noise and Signal in  Data Numbers (DN) were converted to $e^-$ using a conversion gain, $G$. Dark current is negligible at these signal levels. The solid line in Figure~\ref{shot} shows the noise model with a conversion gain of $G=4.2\pm0.2\ e^-$/DN and  $RN=15\pm2\, e^-$. The conversion gain includes a correction for interpixel capacitance of 0.81 \citep{Brown2006}.

\subsection{Light Curve Statistics}

We use the Allan variance \citep{Howe81} as the main measurement statistic for assessing photometric precision. For a light curve $l(t)$, the Allan 
variance $\sigma_l^2$ over a time lag $\tau$ is the second moment of $L^+-L^-$,
\begin{equation}
 \sigma_l^2(\tau) = {1\over 2} \langle(L^+-L^-)^2\rangle ~~~\forall~ \tau \le \frac{T}{2}, 
\end{equation}
where $L^+$ and $L^-$ are integrals (or equivalently, the integrals are
replaced by sums for periodically sampled, discrete lightcurves) over adjoining time intervals of 
duration $\tau$,
$$L^- = {1\over\tau}\int_{t-\tau}^t l(t) dt $$
$$L^+ = \frac{1}{\tau}\int_t^{t+\tau} l(t) dt $$ 
and $T$ is the total duration of the lightcurve. This statistic preferentially selects noise
characteristics on a lag timescale, $\sim \tau$, while rejecting noise on timescales both longer and shorter than $\tau$. For random noise, the variance assumes its usual $1/\tau$ dependence; and in general, $f^{-n}$ noise
has $\sigma_l^2 \propto \tau^{n-1}$. It is illustrative to write the
Fourier domain filter-form of the Allan deviation, as
$$ \sigma_l^2(\tau) = 2\int_0^\infty S_l(f) \sinc^2(\pi \tau f) \sin^2(\pi \tau f) df$$
where $S_l(f)$ is the power spectrum of $l(t)$. This shows that $S_l(f)$ is filtered
by a bandpass filter with a response $H_A(f)$,

$$ |H_A(f)|^2 = 2\frac{\sin^4(\pi \tau f)}{(\pi\tau f)^2},$$
peaking at $f \simeq 1.2/\pi \tau$.

For a given experiment, we calculate an Allan deviation for each star within the detector region of interest. The averaged or global Allan deviation is used as the statistic for the determining how well we did over the ensemble of stars in the experiment and is given as, $\sigma_L(\tau) = \sqrt{\langle \sigma_l^2(\tau)\rangle}, $ where the average $\langle \rangle$ is over the ensemble, after rejection of outliers that show exceptionally poor Allan deviations. About $\sim 3 - 5$ \% of all stars are rejected as outliers with anomalous Allan deviations. In this paper we do not discuss the nature of the the outlier stars other than to say that their spatial distribution is fairly random and that at least some fraction of these stars show poor photometry due to unresponsive pixels located within the aperture.

\begin{figure*}
\epsscale{1}\plotone{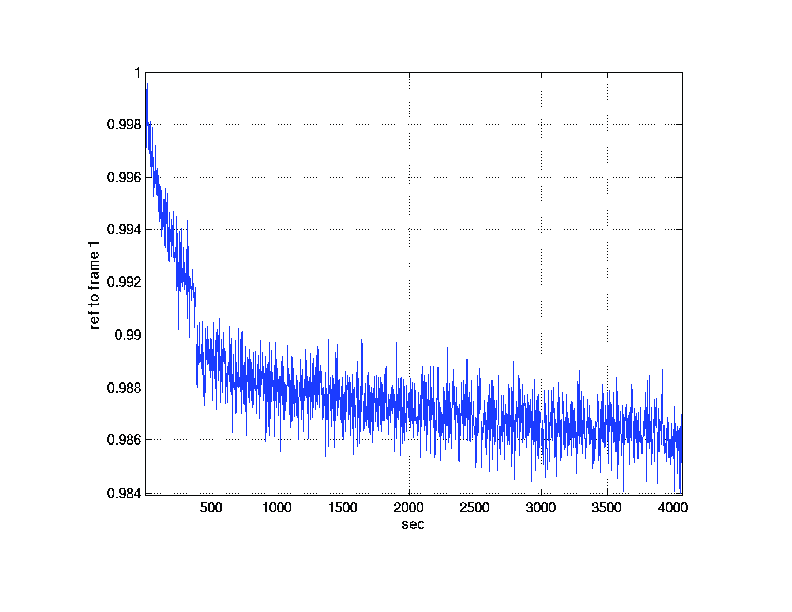} \caption{The light curve of a typical spot for the first hour after turning on the LED shows a rapid drop in LED brightness of $\sim$1.2\%, as expected for diode junction self heating by $\sim$1K, followed by a long term drift of $-0.2$\%/hr.  Short term fluctuations at the level of a few parts per thousand are also evident.\label{LEDDrift}} 
\end{figure*}

When no corrections are applied to light curves, $\sigma_L(\tau)$ quickly departs from the $\tau^{-1/2}$ curve expected 
for Gaussian noise. The primary culprit at this noise level are the $\sim 10^3$ ppm fluctuations of the lamp on timescales 
of minutes (Figure~\ref{LEDDrift}). Because the lamp intensity is not explicitly stabilized by feedback, we rely on common 
mode rejection of lamp intensity variations as described below to reduce this noise source.

\subsection{Measurement Model and Parameter Decorrelation} 

Light curves rapidly depart from the ideal case due to a number of sources of $1/f$ noise. Laboratory lamps vary more rapidly than stars, and lamp fluctuations (10$^3$ ppm) are typically the largest source of non-ideal behavior in our experiments. However, because the lamp intensity is in common to the spot ensemble and because we achieve excellent common-mode rejection, lamp variations can be removed to the required accuracy. The stability of the light curves is also affected by environmental factors such as image wander, changes in the optical point spread function, detector temperature, etc. These changes can be measured and captured in the ``instrument state'' and then be removed from the data. In general, the light curves $\bf y$ (vector-valued for $m$ spots) are related to
the combined source and instrument state, {\bf x}, as

\begin{equation}
 {\bf y} = \bf{F}(\bf x).
\end{equation}

There are various ways in which to define the vector-valued model function {\bf F(x)}. The most sophisticated approach is via a full description of the image formation process if one is in possession of the detailed PSF and the detector response function.
Such an approach may be necessary to obtain good photometric performance in cases when the PSF is poorly sampled and/or if the perturbations are relatively large. In our case, with a well sampled PSF, and for relatively small perturbations to the instrument state, a simpler parameterized form of {\bf F(x)} suffices.  We find that in most instances we can linearize (3) and rewrite it as

\begin{equation}
	{\bf y} = {\bf F}({\bf x_i}) + {\bf K}({\bf x} - {\bf x_i}),
\end{equation}
where {\bf x$_i$} is some initial state, and ${\bf K}$ is a $m\times n$ Jacobian matrix with elements given by
the partial derivatives,
\begin{equation}
K_{i,j} = {\partial F_i({\bf x}) \over \partial x_j}
\end{equation}
relating the photometry for spot $i$ to the instrumental parameter $j$. One of our goals is to quantify the
``sensitivities'' to various parameters; for any parameter $x_j$, this is simply a row average (i.e. equivalent to an average over
the spot ensemble) of the Jacobian, $\langle K_{i,j} \rangle_i$. 

However, in some cases the truncation to first order (4) is insufficient; this
is the case if there is relatively large image motion ($ > 0.1$ pixel) or if lamp fluctuations are large and pixel response non-linearities must be considered. If so, ${\bf F}$ can be expanded to second order in ${\bf x}$. This higher order model is  recommended, for example, for light curves extracted from poorly sampled Spitzer IRAC Bands 1 and 2 images \citep{Knutson2011}, requiring estimation of the Hessian,
\begin{equation}
H_{i,jk} =  {\partial F_i({\bf x}) \over \partial x_j \partial x_k}
\end{equation}

In our experiments the state is well determined by the following set of parameters:

\begin{enumerate}
\item {\color{black} The image centroid or pointing drift, $\delta x$ and $\delta y$, measured in pixels, and usually a small fraction of the pixel pitch. 
In general, we find that for image motions larger than a
tenth of a pixel, $\delta x,\delta y \ge 0.1$ pixels, 
second order terms in $\delta x$ and $\delta y$ are required to model the photometry. With image motions of $\ge 0.3$ pixels the second order model no 
longer suffices.}
\item Changes in the optical point spread function. Focus drifts can lead to significant changes in the PSF and the photometry.
We capture focus with relatively simple parameterization, i.e., we approximate all spots as Gaussians and track the change in the
FWHM ($\delta w_x, \delta w_y$) caused by temporal focus drifts.  However, we have no way to inject controlled amounts of focus error.
\item Changes in the lamp intensity ($\delta L$) are estimated by an average of the lightcurves of an ensemble of stars randomly distributed over the detector ROI. In typical experiments, about 20 stars are need for a reasonable estimate of $\delta L$.
\item Any changes in the detector temperature ($\delta T_{FPA}$). 

\end{enumerate}

Often changes in the state parameters are correlated, being driven by the same underlying mechanism, i.e., the daily thermal cycles between dewar fills. On orbit this could be the solar thermal forcing of a spacecraft. Given these correlations it is convenient to transform the state ${\bf x}$ to ${\bf x'}$, with the latter having a minimum set of uncorrelated or orthogonal parameters by using a principal components analysis (PCA).  PCA is an orthogonal linear transformation that maps a set of observations of possibly correlated vectors into a set of values of orthogonal vectors, called the principal components. By definition the first principal component has the highest variance, i.e. it accounts for as much of the variability in the observed data as possible. Subsequent principal components are ordered by variance but under the constraint of orthogonality to the preceding components.
Our analysis proceeded according to following steps:

\begin{itemize}
\item Let the light curve for each spot consist of $p$ measurements; each measurement yields an estimate of the state parameters. Convert the 
$p \times 1$ vectors of state parameters $x_j$ to centered, rescaled versions via the transformation
$\hat{x_j} = (x_j-\bar{x_j})/\sigma_{x{_j}}$, where $\bar{x_j}$ and $\sigma_{x{_j}}$ are the mean and standard deviation. 

\item Construct the $p \times n$ state matrix ${\bf X}$ for each spot using a set of $n$ state parameter vectors. Rows of ${\bf X}$ correspond to the observations, and columns to the parameters.

\item Use the IDL routine PCOMP(${\bf X}$) \footnotemark[3] to perform principal components analysis on ${\bf X}$ and return the principal component coefficients. 
 
\item We generate the new state, $\bf {\hat{X}}$, using a truncated set of $m$ principal component vectors ($m \le n$, and $m$ is typically 3-4). 
The normalized light curve ${\bf Z}$ is described by  ${\bf Z = \hat{X} K + e}$, where ${\bf e}$ is an error vector. 
Minimizing the cost, ${\bf e^Te}$, with respect to ${\bf K}$, yields a least-squares solution for the sensitivities
${\bf K = (\hat{X}^T\hat{X})^{-1}\hat{X}^T{Z}}$. Repeated tests show the error ${\bf e}$ to be nearly zero mean Gaussian, making the least squares estimate of {\bf K} equivalent to a maximum likelihood estimate, and conferring upon
it a direct probabilistic interpretation.

\item Compute the model light curves, ${\bf \hat{X}K}$. The error residuals, ${\bf e = Z - \hat{X}K}$ are used to compute the the Allan deviations. Allan deviation curves labeled as ``decorrelated" have been corrected for systematics using the above scheme.

\end{itemize}

\section{Results\label{results}}

\begin{figure*}
\epsscale{1}\plottwo{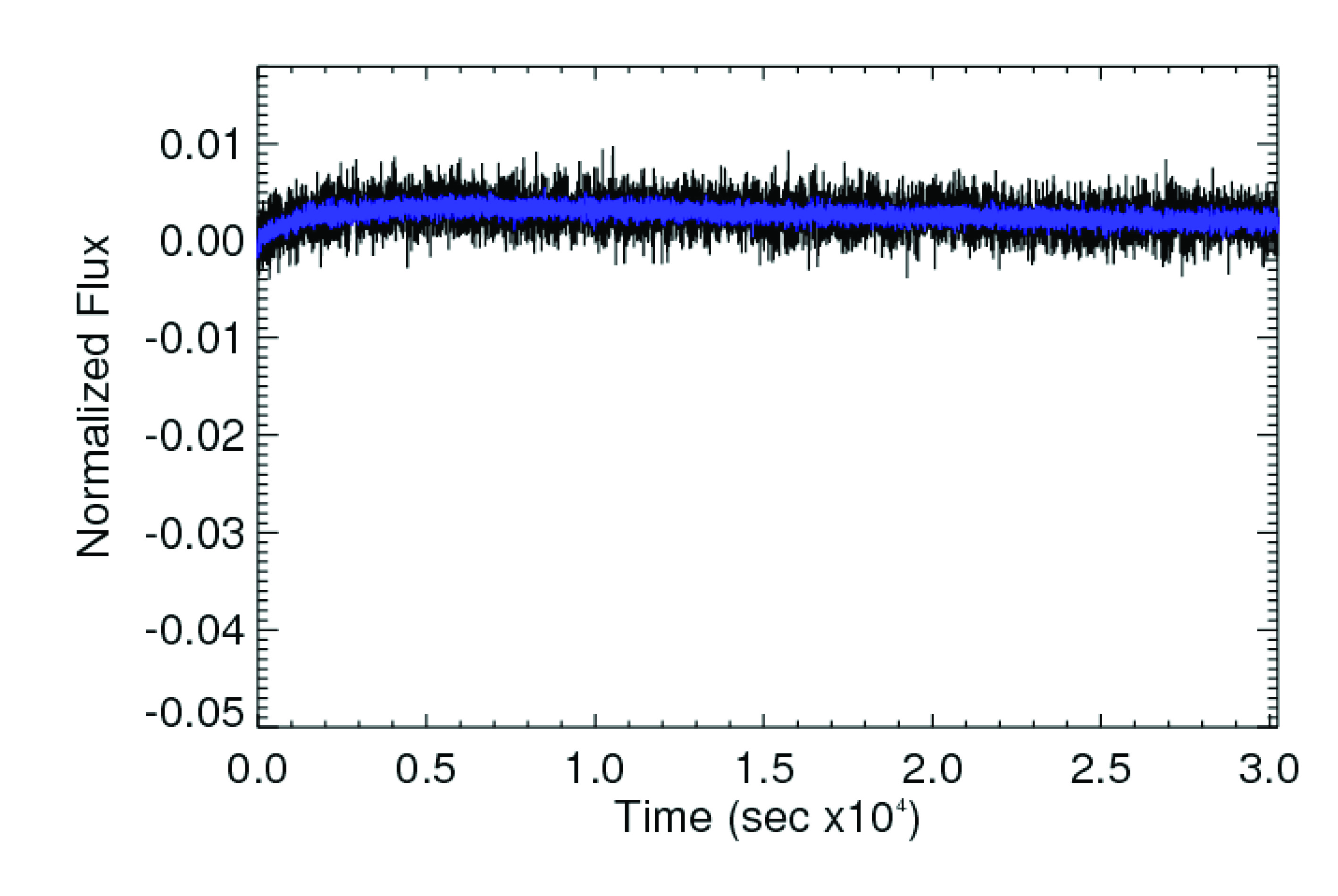}{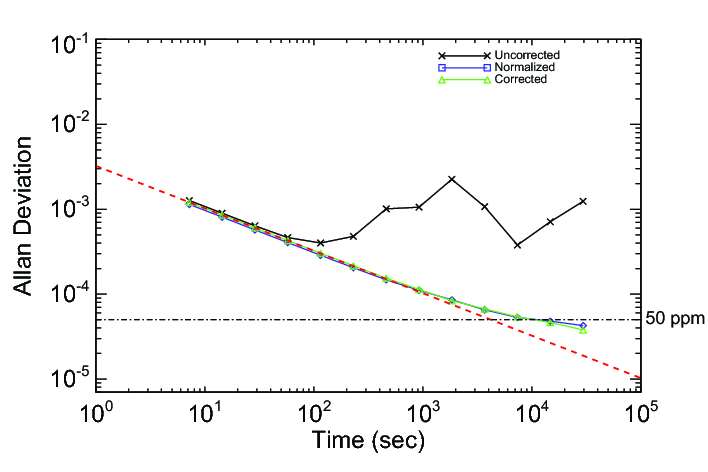}
\caption{a,left) A single spot light curve ($I_i/\bar{I}-1$) over an 8 hr interval for nominally constant lamp intensity shows variation at the level of a few tenths of a percent (black curve). The PCA model is shown in magenta. b, right) The Allan deviation of the light curve (black trace), inclusive of lamp variations, hits a floor at the 0.1\% (or 1000 ppm) level. After PCA decorrelation the noise level obeys a $\tau^{-0.5}$ relationship until reaching a floor of roughly 50 ppm after 4.2 hours for single spot before (blue curve) and after bad pixel rejection (green curve)}.\label{AV1}
\end{figure*}

\begin{figure*}
\epsscale{1}\plotone{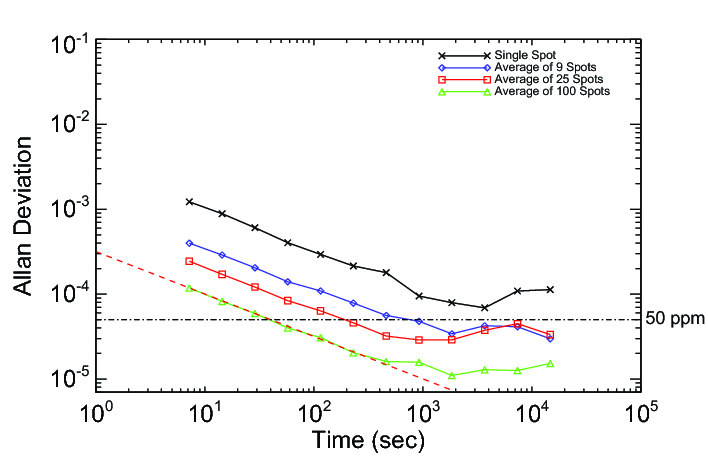} \caption{ Allan deviations of decorrelated light curves, based on data presented in Figure ~\ref{AV1}a made by co-adding multiple spots (from top to bottom: $N = $ 1, 9, 25, and 100 spots).  The noise level decreases as $\sqrt{N}$ at short lags where the statistics are determined by photon noise. Even at longer lags, where
correlated noise dominates single spot Allan deviations, the improvement is $\sim \sqrt {N}$, suggesting that the systematics are mostly uncorrelated {\it between}
spots. The dashed line shows that when sufficient flux is included in the measurement, i.e. the 100 spot line, the noise maintains 
the photon noise limited $\tau^{-0.5}$ relationship until reaching a floor of roughly $\sim 10 - 20$ ppm.} \label{Nspots}

\end{figure*}

\begin{figure*}
\epsscale{1}\plottwo{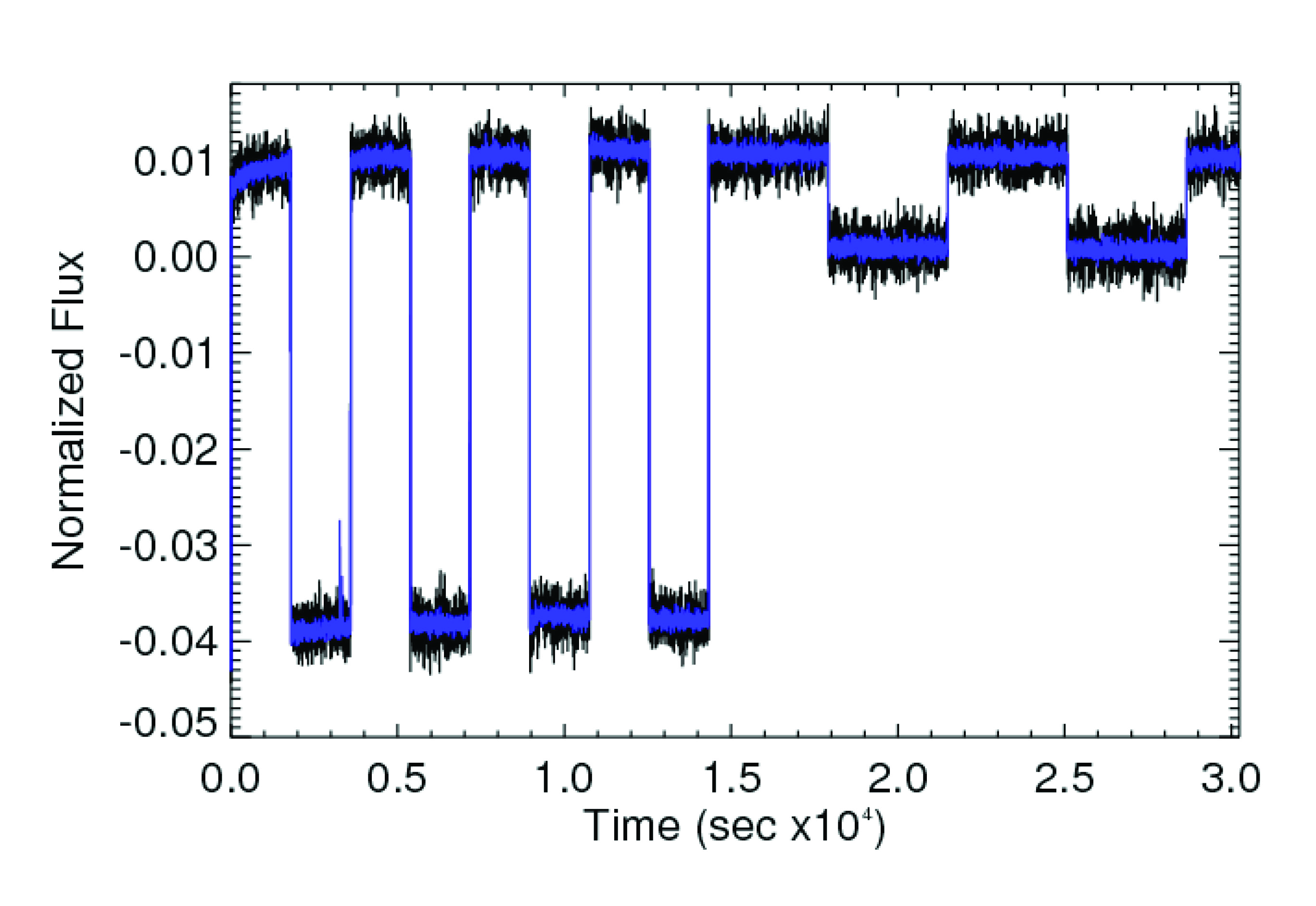}{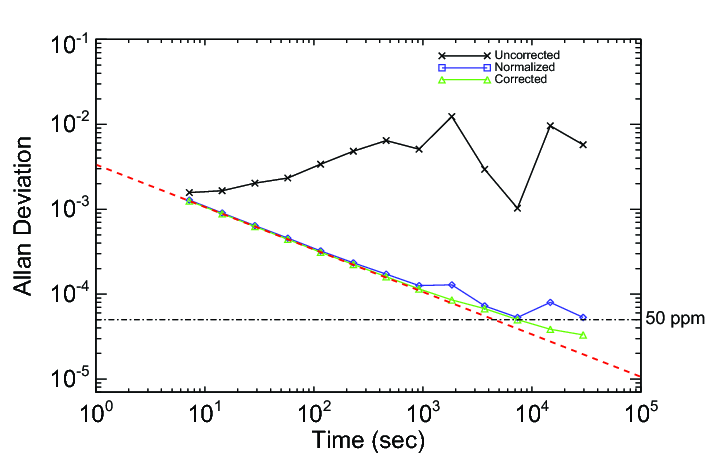} \caption{left)  The light curve of a single spot following commanded changes in lamp intensity. right) The 
averaging characteristics of unnormalized data are in black with 'x' markers. After normalization by the ensemble average, the noise floor drops to $<$100 ppm (blue curve, diamonds). After full PCA decorrelation, the noise level follows a $\tau^{-0.5}$ relationship until reaching a floor of roughly $<$50 ppm for a single spot (green curve, triangles). The common mode rejection for lamp variations is better than 100:1.\label{CommandVar}}
\end{figure*}

In the simplest experiments, we held the experimental setup constant for approximately 24 hr and either kept  the lamp brightness constant or changed the lamp brightness in discrete steps. Figure~\ref{AV1}a  shows the  intensity of a single spot before correction for variations in lamp brightness. 
The Allan deviation as a function of lag for that single spot (Figure~\ref{AV1}b),  inclusive of lamp variations, hits a floor at the 0.1\% (or 1000 ppm) level.  The decorrelation using PCA (with or without clipping of a few bad pixels) greatly reduces the effect of lamp variations and other noise sources,  achieving a noise floor of $\sim$50 ppm for a single spot. Data sets from different experimental runs shown in Figures~\ref{Nspots}-\ref{temperature} show single spot noise floors between $\sim$30-50 ppm.  

Figure~\ref{Nspots} shows the result of first averaging the light curves of multiple spots together and then computing the Allan deviation of the averaged light curve. At short lag times, the initial noise level decreases as expected for photon-noise limited measurements: the noise is inversely proportional to the square root of the number of spots, i.e. proportional to the square root of total number of photons. At longer lag times, the limiting noise floor  also drops as the square root of number of spots, i.e. as the square root of number of independent, uncorrelated measurements. This result implies that averaging over multiple spots  can reduce other temporally-correlated noise sources down to the ultimate noise floor  seen in these experiments  of  10-20 ppm. Figure ~\ref{CommandVar} shows that even large, commanded excursions in lamp brightness can be corrected for via ensemble averaging and PCA decorrelation. Overall, these data show that by using decorrelation techniques and suitable  averaging of independent samples,  the H2RG detector can achieve  very high precision differential photometry with  noise values as low as  $\simlt$20 ppm.

\vskip 0.1 in

\subsection{Pointing Rejection}

In general, the measured count rate for a source represents a convolution of the detector spatial response with the pattern of optical illumination over a predefined aperture. The illumination pattern is the broadband PSF in the case of a survey experiment, or a narrow spectral extraction region in a spectrograph. Because of the convolution the coupling of pointing changes and/or PSF variations with inter-pixel or intra-pixel responsivity variations within the detector changes the measured signal. While a transit survey experiment can correct for gross drifts in detector/electronics response using stellar ensemble averages, changes in the measured brightness of individual stars due to small pointing shifts must be corrected using decorrelation techniques. These techniques have been pioneered using Spitzer and HST data  \citep{Knutson2007, Swain2008}.

The response within a pixel can have substantial structure due to declining detection efficiency at its edges, diffusion of charge into neighboring pixels and electrical crosstalk between pixels. Sensitivity variations across a pixel reduce photometric accuracy in images even where the PSF is fairly well sampled. The intrapixel variations in the PACE type detectors flown on NICMOS have been found to be quite large \citep{Finger2000} and clearly affect the lightcurve fidelity obtained during NICMOS transit photometry, despite the large (5-pixel FWHM) defocus in the NIC3 camera. 

\begin{figure*}
\epsscale{1}\plottwo{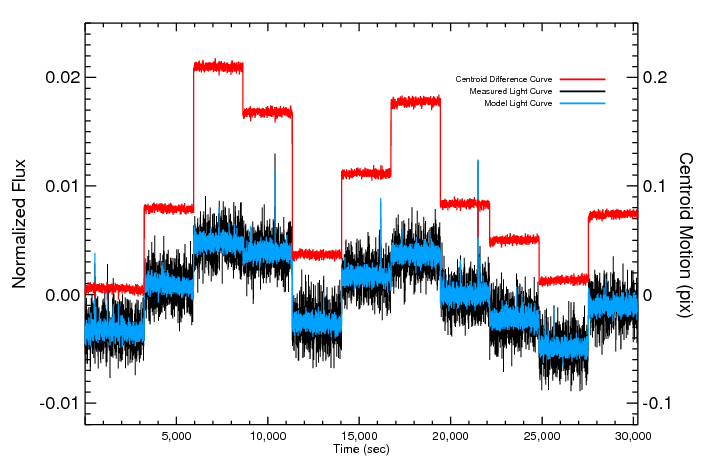}{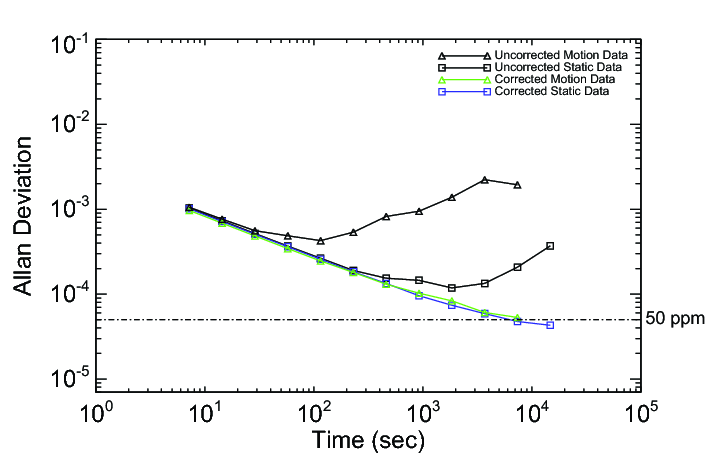}\caption{ left) Light curve of a single spot in response to commanded image motion. The large steps (right hand axis in pixel units) represent centroid motions of up to 0.25 pixel. The thick black curve shows the resultant variations in measured signal intensity, up to almost 1\%. The lighter colored (blue) curve shows the PCA model fitted to the light curve data. right) The curves marked with squares denote data with no motion while the curves with triangles denote data with position offsets. The two lower curves were corrected using PCA analysis. After full PCA decorrelation the noise level follows a $\tau^{-0.5}$ relationship until reaching a floor of roughly 30 ppm for a single spot. \label{position1}} 
\end{figure*}

These intra- and inter-pixel responsivity variations make pointing jitter and wander amongst the largest terms in the photometry error budget in a transit experiments using older detectors. Improvements to quantum efficiency of HgCdTe detectors in recent years have resulted in most photons being detected if they are not reflected. Even though lateral charge diffusion results in a variation in response of a given pixel depending on the position of the stellar image the integrated response remains almost constant. We conducted an experiment to determine the degree to which pointing affects photometric precision, and the degree to which decorrelation analysis can be used to remove these effects from the lightcurves. Controlled pointing offsets were introduced by rotating the dewar and causing a 1-D shift in spot positions ($\S$2, Figure~\ref{rotation}). A series of dewar rotations produced up to 0.25 pixel steps in the X locations of the spots and much smaller shifts in the Y-locations. After determining the new spot centroids, we applied the PCA decorrelation technique and were able to reduce the noise floor to well under 30 ppm as demonstrated in Figure~\ref{position1}. We conclude that in a configuration with $\sim$ 2 pixel radius spots, photometry of $<$50 ppm can be achieved using PCA correction even in the presence of pointing offsets of up to 0.25 pixel.

\vskip 0.1truein 
\vskip 0.1truein

\subsection{Temperature Dependence}

\begin{figure*}
\epsscale{1}\plottwo{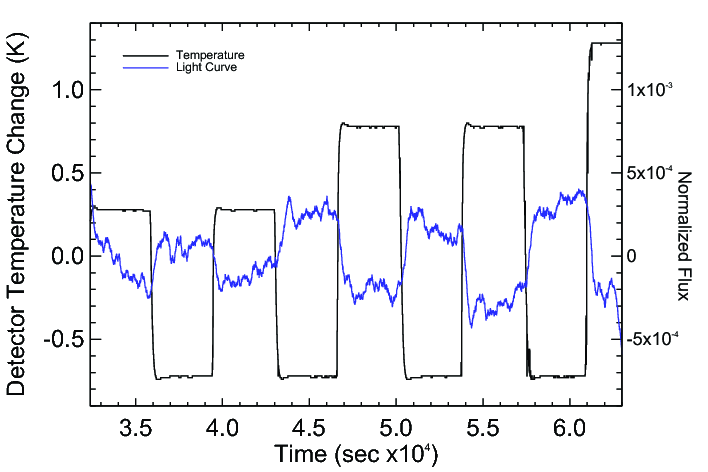}{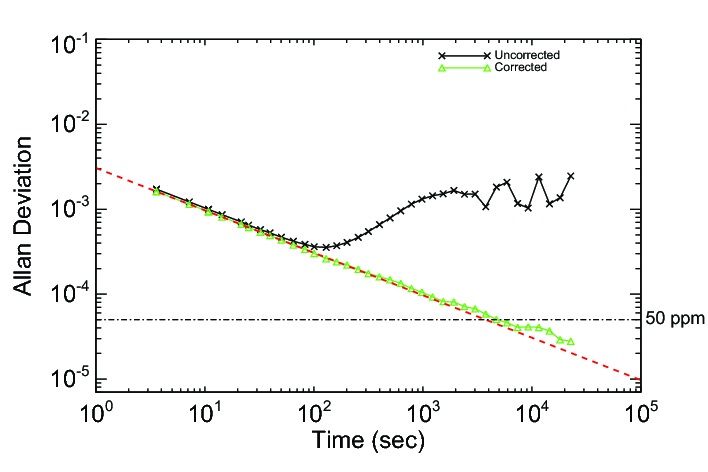} \caption{left) A single spot light curve responding to commanded steps in the detector temperature in black, right) An Allan deviation plot showing unnormalized data (black curve) at few percent level. After normalization by the ensemble average, the noise floor drops to $<$100 ppm (blue curve). After full PCA decorrelation the noise level follows a $\tau^{-0.5}$ relationship until reaching a floor of roughly $<$30 ppm. \label{temperature}} 
\end{figure*}

Thermal fluctuations can produce changes in detector response. Fortunately, the temperature dependence of detector responsivity is relatively weak in modern H2RG arrays compared with variation in the quantum efficiency in the NICMOS 3 devices used on HST. We conducted a simple experiment to determine the degree to which the detector temperature must be controlled to avoid impacting precision photometric requirements. In our set-up a temperature controller maintains the focal plane at 140 K with a precision of $\pm$5 mK. An 8.5 hr long photometric image sequence was acquired during which the temperature of the focal plane was modulated by changing the controller set-point temperature by up to 2 K. The temperature steps are shown in Figure~\ref{temperature} along with the change in the average signal. The figure also shows the Allan deviation for the uncorrected and corrected (decorrelated) data. The decorrelation technique can correct for relatively large swings in detector temperature without a significant impact on the final photometric precision. 

The average temperature coefficient of the intensity variation found by averaging the lightcurves of many stars is $\sim250$ ppm/K. The combination of temperature control at the 0.1 K level and decorrelation analysis will reduce the effect of temperature fluctuations to acceptable levels. A flight instrument would use a variety of techniques to maintain detector temperature including heaters at the focal plane level (0.05 K) as well as using thin metal traces built into the H2RG multiplexor that can monitor and control the package temperature directly at the few milli-K level. Finally, detector reference pixels can be used to suppress offsets due to residual bias voltage and temperature drifts.

\section{Discussion}

\subsection{Applicability to Other Near-IR Focal Planes}

How applicable are these results to Teledyne HxRGs with different cutoff wavelengths?    H2RG-103, used in these tests, is from one of the first successful batches of 1.7 $\mu$m cutoff HgCdTe detectors made for the SNAP program in 2006. Substrate removal had been mastered, delivering what is believed to be near 100\% internal quantum efficiency: almost all photons passing through the anti-reflection coating produce signal electrons.  A corollary of this high internal QE, shared by devices with other cutoff wavelengths, is that intrapixel response variation is low since there is no charge loss mechanism to modulate sensitivity.

Barron et al. (2007) performed intra-pixel response measurements on the near-identical 1.7 $\mu$m cutoff H2RG-102 from the same batch, and found that, while lateral charge diffusion redistributes charge between neighboring pixels, the integrated flux is independent of the position of sub pixel size spots at the $<$2\% rms level.    Hardy et al. (2008) reached the same conclusion after performing similar tests over a wider area on JWST-like H1RGs\footnote{H1RG or HAWAII-1RG: HgCdTe Astronomical Wide Area Infrared Imager} \citep{Beletic2008} with 5 $\mu$m cutoff and found that process dependent features, such as surface particulates and sub pixel width "scratches" dominated the response.

The higher bandgap of the 1.7 $\mu$m detector presents special manufacturing challenges that result in higher read noise and persistence.  While operating temperature can be higher for a given dark current, the dark current floor is similar to the longer cutoff devices.    For these reasons we believe that the 1.7 $\mu$m devices represent a worst-case scenario, and that the results in this paper are probably extensible to longer cutoff devices.  It is often the case that greater differences in performance are observed due to process variations than due to the cutoff itself, but the better devices at any cutoff wavelengths are expected to reach the levels of performance reported here.

Thus, we maintain that our  results  showing 1.7 $\mu$m cutoff H2RG devices  are capable of $<$ 50 ppm precision photometry are directly applicable to longer wavelength devices such as JWST's  5 $\mu$m arrays. Our existing experimental setup could be used to probe the behavior of 2-2.5 $\mu$m cutoff material with only slight modifications. However, operation out to 5 $\mu$m would require a new experiment designed with much more attention to reduced backgrounds and lower temperature operation.

\subsection{A Near-IR Transit Survey}

\begin{figure*}
\epsscale{1}\plotone{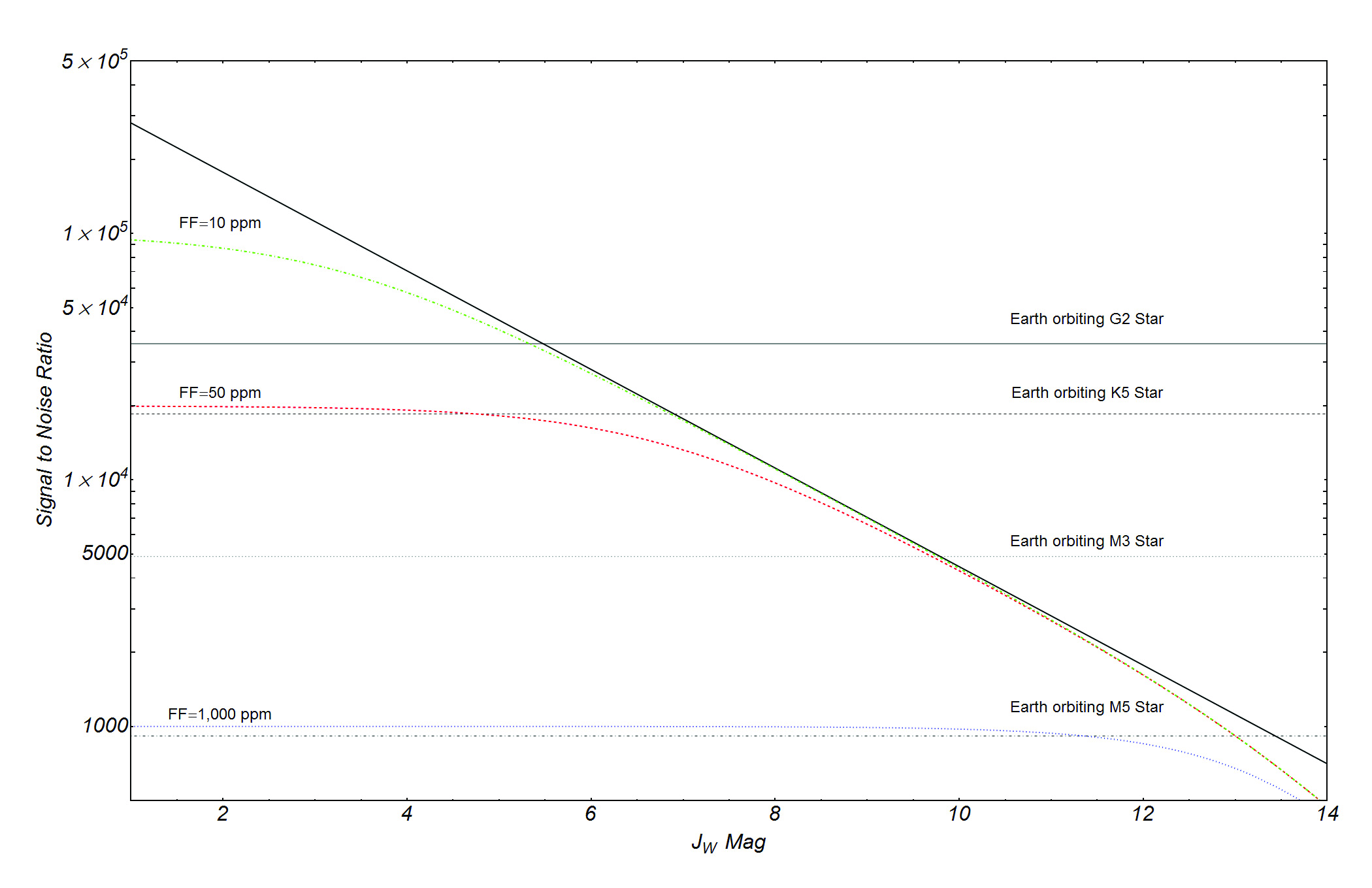}
\caption{Signal to Noise ratio (SNR) for stars as a function of  broad band (0.66-1.70  $\micron$, denoted J$_W$) magnitude after a 900 sec integration typical of an observation in Low Earth Orbit. Observations are assumed to use a camera with 12.5 cm aperture using a 1.7 $\mu$m cutoff H2RG detector as discussed herein. The noise model includes photon noise from the star and the sky as well as a noise floor set by the limiting detector performance (the ``Flat Field" limit, $FF$) in ppm. The solid black line denotes the noise limit as set only by stellar photon noise. At  magnitudes fainter than J$_W\sim$14 mag sky background reduces the SNR while for bright stars the residual noise floor dominates the SNR. The effects of stellar jitter and background confusion are neglected here but are included in the Monte Carlo simulations. The horizontal lines show the SNR needed to detect a  1 $R_\oplus$  planet transiting a G2, K5, M3 star in a single 900 sec observation at the 3$\sigma$ level. \label{ELEKTRA}} 
\end{figure*}

We put this detector performance into context by considering  observations with a small near-IR camera (12.5 cm diameter) observing in a wide ``J$_W$'' band extending from 0.66 to 1.65 $\micron$ where we have taken typical values for the overall optical and detector quantum efficiency (total of 50\%) and sky brightness in space (0.15 MJy sr$^{-1}$). In Figure~\ref{ELEKTRA} the solid black curve shows the idealized case where the SNR varies simply as the square root of the stellar photon flux. Three additional curves show realistic cases taking into account the photon noise from sky background and 3 different values for the noise floor (10, 50 and 1,000 ppm). For the bright stars considered here, detector read noise and dark current are negligible sources of noise. For the brightest stars, the SNR curves flatten out due to the uncorrected detector responsivity changes, i.e. the flat field ``noise floor''. At faint magnitudes, photon noise from the sky background reduces the SNR relative to the stellar photon noise limit. The 1,000 ppm limit represents the best values that have been obtained in ground based surveys of very large numbers of stars ($\S1)$. Horizontal lines denote the SNR levels needed to detect the transit of a 1 R$_\oplus$ planet orbiting G2, K5, and M3 stars at the 3$\sigma$ level. The noise floor demonstrated in this paper, $<$50  ppm is adequate to detect at Earth transiting a K star at 10-20 pc in a single 900 sec observation where 900 seconds is a typical observation time for a spacecraft in Low Earth Orbit.  The fact that the noise in the average of independent measurements continues to decrease to $\sim$10-20 ppm ($\S$4) implies that weaker transit signals can be extracted from the light curves with suitable processing of multiple transits.   Typical requirements for a transit survey demand a minimum of 3 transits and a total SNR=7 for a confirmed detection.

\begin{figure*}
\epsscale{1}\plotone{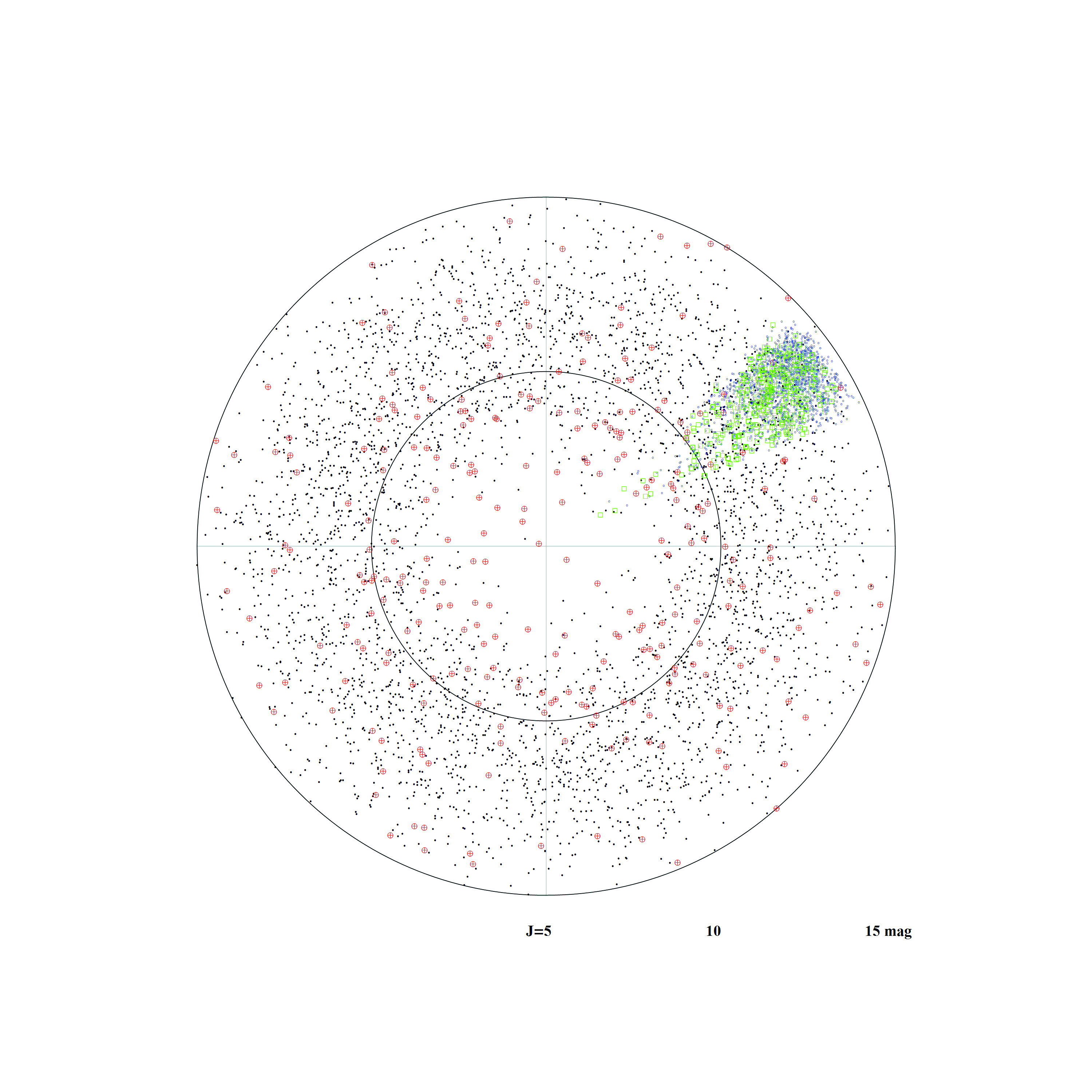}
\caption{A Monte Carlo simulation suggests that an all sky survey would find thousands of gas giant planets and hundreds of
rocky planets, many of which would occupy the temperate habitable zones of their late-type host stars. In this polar diagram, the radial coordinate represents a J magnitude and the polar angle represents a simulated sky coordinate. Gas giants from the simulation are shown as small filled circles while rocky planets are shown as red $\oplus$s. For comparison, confirmed Kepler planets and Kepler candidates are shown with their J (2MASS) magnitudes \citep{Borucki2011b, Batalha2012}. Kepler's gas giant planets are shown as blue circles and rocky planets as red squares.  \label{Monte}} 
\end{figure*}

The yield of a survey depends on many instrumental and survey strategy parameters, e.g. number of cameras, survey cadence and duration, mission length, etc.,  and is beyond the scope of this article. We have developed a Monte Carlo simulation using modest extrapolations of the distributions of the size and orbital location of planets  derived from  radial velocity and transit surveys \citep{Howard2011, Borucki2011b}. The simulation, which will be described elsewhere, conservatively estimates that for a sample of $2 \times 10^6$ stars each hosting exactly one planet and spanning spectral types from F0 to M4 ($J_W<14$ mag) and M5-M8 ($J_W<15$ mag), a 1.5 yr near-IR survey using 6 cameras on a spacecraft in a low earth orbit would detect over $\sim$200  rocky planets ($\simlt 2\ R_\oplus$)  (Figure~\ref{Monte}) of which at least $\sim$50 would be located in the habitable zones of their K and M star hosts. The time any star is monitored varies with ecliptic latitude, $\beta$, and for the particular configuration simulated here we assumed 75 days per year for a star at $\beta = 30^{\circ}$. The survey would also identify a few thousand gas and ice giant planets covering a broad range of equilibrium temperatures. Some 300 planets would orbit stars brighter than J$_W=9$ mag and be suitable for confirmation and mass determination via radial velocity measurements and spectroscopy  for the characterization of  planetary atmospheres. A survey conducted in a more distant orbit from the Earth, e.g. L2, would offer a greatly improved observing cadence  resulting a more complete survey and better characterized transit curves.

Due to the fundamental selection biases of transit surveys of modest duration, most transiting planets will be found in  short period orbits of a few months or less, but only for late type stars does this preferred orbital location coincide with the habitable zone. A near-IR survey ($\S1$, Figure~\ref{NIR}) would find   3-7 times more rocky planets orbiting K and M stars than an otherwise identical visible light survey, a critical difference for the identification of  planets with temperate atmospheres ($250\,\mathrm{K}<T_{eff}<350$ K) which are of great interest for spectroscopic follow-up. The laboratory experiments described in this paper demonstrate that the performance of modern near-IR detectors is adequate for such a survey and for subsequent follow-up observations. 

\subsection{Future Work}

Experiments with the existing dewar configuration are currently underway to correlate spatial variations detector performance (e.g. dark current, responsivity, and linearity) with photometric performance, both to identify whether calibrations can improve upon the photometric performance obtained through PCA alone and to establish criteria for excluding "bad pixels". These criteria will assist in developing metrics for detector screening and grading for possible future missions. Future experiments in this series will include 1) an examination of the effect on photometric precision of varying the number of pixels within the stellar image which is directly relevant to JWST/NIRCam observations using weak lenses to defocus NIRCam images and 2) simulations of the extraction of transit spectra over a wide range of spectral line properties, e.g. line depth and line width. These results will be relevant to photometric, grism and spectroscopic modes of JWST as well as to dedicated transit spectroscopy missions presently under consideration in the US and Europe.

\acknowledgments

The research described in this publication was carried out in part at the Jet Propulsion Laboratory, California Institute of Technology, under a contract with the National Aeronautics and Space Administration. Support from the JPL Research and Technology Development program as well as the from the Near-IR Camera (NIRCam) instrument on James Webb Space Telescope are gratefully acknowledged.  We are grateful to Thomas Greene and Jim Beletic for useful discussions. The HR2G detector used for these experiments was provided by the Lawrence Berkeley National Laboratory.


\begin{thebibliography}{}
%\expandafter\ifx\csname natexlab\endcsname\relax\def\natexlab#1{#1}\fi
\bibitem[Baglin et al.(2009)]{Baglin2009} Baglin, A., Auvergne, 
M., Barge, P., et al.\ 2009, IAU Symposium, 253, 71. 
\bibitem[Bakos et al.(2009)]{Bakos2009} Bakos, G.~{\'A}., Noyes, 
R.~W., Kov{\'a}cs, G., et al.\ 2009, IAU Symposium, 253, 21. 
\bibitem[Barman(2007)]{Barman2007} Barman, T.\ 2007, \apjl, 661, 
L191. 
\bibitem[Barman(2008)]{Barman2008} Barman, T.~S.\ 2008, \apjl, 
676, L61.
\bibitem[Barron et al.(2007)]{Barron2007} Barron, N., Borysow, M., 
Beyerlein, K., et al.\ 2007, \pasp, 119, 466. 
\bibitem[Basri et al.(2010)]{Basri2010} Basri, G., Walkowicz, 
L.~M., Batalha, N., et al.\ 2010, \apjl, 713, L155. 

\bibitem[Batalha et al. (2012)]{Batalha2012} Batalha, N.  et al. 2012, \apjs, in press. 
\bibitem[Bean et al.(2010)]{Bean2010} Bean, J.~L., Seifahrt, A., 
Hartman, H., et al.\ 2010, \apj, 713, 410. 
\bibitem[Beaulieu et al.(2011)]{Beaulieu2011} Beaulieu, J.-P., 
Tinetti, G., Kipping, D.~M., et al.\ 2011, \apj, 731, 16. 
\bibitem[Beichman et al.(2009)]{Beichman2009} Beichman, C.~A., 
Greene, T., \& Krist, J.\ 2009, IAU Symposium, 253, 319. 
\bibitem[Beletic et al.(2008)]{Beletic2008} Beletic, J.~W., Blank, 
R., Gulbransen, D., et al.\ 2008, \procspie, 7021.
\bibitem[Berta et al.(2012)]{Berta2011} Berta, Z.~K.,
Charbonneau, D., D{\'e}sert, J.-M., et al.\ 2012, \apj, 747, 35

\bibitem[Brown et al.\ (2006)]{Brown2006} Brown, M., Schubnell,M. \& Tarle, G., 2006, \pasp,118,1443.
\bibitem[Biesiadzinski et al.(2011)]{Bies2011} Biesiadzinski, 
T., Lorenzon, W., Newman, R., et al.\ 2011, \pasp, 123, 958. 
\bibitem[Borucki \& Koch(2011)]{Borucki2011} Borucki, W.~J., \& Koch, D.~G.\ 2011, IAU Symposium, 276, 34.

\bibitem[Borucki et al.(2011)]{Borucki2011b} Borucki, W.~J., Koch, D.~G., Basri, G., et al.\ 2011, \apj, 728, 117 
\bibitem[Bryden et al.(2011)]{Bryden2011} Bryden, G., Stauffer, J., Ciardi, D.~R., 
\& NStED Science Team 2011, Bulletin of the American Astronomical Society, 43, \#140.20. 
\bibitem[Charbonneau et al.(2009)]{Charb2009} Charbonneau, D., Berta, Z.~K., Irwin, J., et al.\ 2009, \nat, 462, 891.
\bibitem[Ciardi et al.(2011)]{Ciardi2011} Ciardi, D.~R., von Braun, K., Bryden, G., et al.\ 2011, \aj, 141, 108. 
\bibitem[Clampin(2010)]{Clampin2010} Clampin, M.\ 2010, Pathways Towards Habitable Planets, 430, 167. 
\bibitem[Claudi(2010)]{Claudi2010} Claudi, R.\ 2010, \apss, 328, 319. 
\bibitem[Deming et al.(2006)]{Deming2006} Deming, D., Harrington, J., Seager, S., \& Richardson, L.~J.\ 2006, \apj, 644, 560 
\bibitem[Deming et al.(2009)]{Deming2009} Deming, D., Seager, S., 
Winn, J., et al.\ 2009, \pasp, 121, 952.
\bibitem[Demory et al. (2012)]{2012ApJ...751L..28D} Demory, B.-O., Gillon, M., Seager, S., et al. \ 2012, \apjl, 751, L28
\bibitem[D{\'e}sert et al.(2009)]{Desert2009} D{\'e}sert, J.-M., 
Lecavelier des Etangs, A., H{\'e}brard, G., et al.\ 2009, \apj, 699, 478.
\bibitem[D{\'e}sert et al.(2011)]{Desert2011} D{\'e}sert, J.-M., Sing, D., Vidal-Madjar, A., et al.\ 2011, \aap, 526, A12  
\bibitem[Finger et al.(2000)]{Finger2000} Finger, G., Mehrgan, H., 
Meyer, M., et al.\ 2000, \procspie, 4008, 1280.
\bibitem[Frasca et al.(2009)]{Frasca2009} Frasca, A., Covino, E., Spezzi, L., et al.  2009, \aap, 508, 1313 
\bibitem[Gaudi (2007)]{Gaudi2007} Gaudi, B. S. 2007, ASPC, 366, 273.
\bibitem[Gibson et al.(2011)]{Gibson2011} Gibson, N.~P., Pont, F., 
\& Aigrain, S.\ 2011, \mnras, 411, 2199.
\bibitem[Gould et al.(2003)]{Gould2003} Gould, A., Pepper, J., 
\& DePoy, D.~L.\ 2003, \apj, 594, 533. 
\bibitem[Hardy et al.(2008)]{Hardy2008} Hardy, T., Baril, M.~R., 
Pazder, J., \& Stilburn, J.~S.\ 2008, \procspie, 7021.
\bibitem[Hauschildt et al.(1999)]{haus99} Hauschildt, P.~H., Allard, F., Ferguson, J., Baron, E., 
\& Alexander, D.~R.\ 1999, \apj, 525, 871. 
\bibitem[Howard et al.(2011)]{Howard2011} Howard, A.~W., Marcy, 
G.~W., Bryson, S.~T., et al.\ 2011, arXiv:1103.2541, \apj, in press.
\bibitem[Howe et al. (1981)]{Howe81} Howe, D. A., Allan, D. W., \& Barnes, J. A.\ 1981, {\it Proceedings of the 35th Annual Frequency Control Symposium,} 1981. (BIN: 554)
\bibitem[Irwin et al. (2010)]{Irwin2010}Irwin et al. 2010, \apj, 718, 1353.
\bibitem[Jenkins et al.(2011a)]{Jenkins2011a} Jenkins, J.~S., Murgas, F., Rojo, P., et al.\ 2011, \aap, 531, A8.
\bibitem[Jenkins et al.(2011b)]{Jenkins2011b} Jenkins, J.~M., Dunham, E.~W., Argabright, V.~S., et al.\ 2011, American Astronomical Society, ESS meeting \#2, \#19.14, 2, 1914. 
\bibitem[Johnson et al.(2009)]{Johnson2009} Johnson, J.~A., Winn, 
J.~N., Cabrera, N.~E., \& Carter, J.~A.\ 2009, \apjl, 692, L100. 
\bibitem[Knutson et al.(2007)]{Knutson2007} Knutson, H.~A., 
Charbonneau, D., Allen, L.~E., et al.\ 2007, \nat, 447, 18.
\bibitem[Knutson et al.(2011)]{Knutson2011} Knutson, H.~A., 
Madhusudhan, N., Cowan, N.~B., et al.\ 2011, \apj, 735, 27. 
\bibitem[Pollacco et al.(2006)]{Pollacco2006} Pollacco, D.~L., 
Skillen, I., Collier Cameron, A., et al.\ 2006, \pasp, 118, 1407. 
\bibitem[Plavchan et al.(2011)]{Plavchan2011} Plavchan, P., Anglada, 
G., \& NIR RV collaboration 2011, American Astronomical Society, ESS meeting \#2, \#20.01, 2, 2001 
\bibitem[Reiners et al.(2010)]{Reiners2010} Reiners, A., Bean, 
J.~L., Huber, K.~F., et al.\ 2010, \apj, 710, 432. 
\bibitem[Ricker et al.(2010)]{Ricker2010} Ricker, G.~R., Latham, 
D.~W., Vanderspek, R.~K., et al.\ 2010, Bulletin of the American 
Astronomical Society, 42, \#450.06. 
\bibitem[Smith et al.(2008)]{Smith2008} Smith, R.~M., Zavodny, 
M., Rahmer, G., \& Bonati, M.\ 2008, \procspie, 7021.
\bibitem[Stevenson et al.(2010)]{Stevenson2010} Stevenson, K.~B., 
Harrington, J., Nymeyer, S., et al.\ 2010, \nat, 464, 1161. 
\bibitem[Swain et al.(2008)]{Swain2008} Swain, M.~R., Bouwman, 
J., Akeson, R.~L., Lawler, S., \& Beichman, C.~A.\ 2008, \apj, 674, 482. 
\bibitem[Swain et al.(2009)]{Swain2009} Swain, M.~R., Vasisht, 
G., Tinetti, G., et al.\ 2009, \apjl, 690, L114. 
\bibitem[Tessenyi et al. (2012)]{Tessenyi2012} Tessenyi, M., 
Ollivier, M., Tinetti, G., et al.\ 2012, \apj, 746, 45.  
\bibitem[Tinetti et al. (2011)]{Tinetti2011} Tinetti, G., et al. 2011, arXiv:1112.2728, accepted for publication in Experimental Astronomy
\bibitem[Tinetti et al.(2007)]{Tinetti2007} Tinetti, G., 
Vidal-Madjar, A., Liang, M.-C., et al.\ 2007, \nat, 448, 169.
\end{thebibliography}
\end{document}